\documentclass{article}
\usepackage{spconf,amsmath,graphicx}
\usepackage[table]{xcolor}
\usepackage{multirow}
\usepackage{cite}
\usepackage{booktabs}
\usepackage{hyperref}
\usepackage{url}
\title{SVDD 2024: The Inaugural Singing Voice Deepfake Detection Challenge}
\name{You Zhang$^{1}$, Yongyi Zang$^{1}$, Jiatong Shi$^{2}$, Ryuichi Yamamoto$^{3}$, Tomoki Toda$^3$, Zhiyao Duan$^1$\thanks{This work is supported in part by a New York State Center of Excellence in Data Science award, National Institute of Justice (NIJ) Graduate Research Fellowship Award 15PNIJ-23-GG-01933-RESS, National Science Foundation (NSF) grants 1846184 and 2222129, synergistic activities funded by NSF grant DGE-1922591, and JST CREST JPMJCR19A3, Japan.}}

\address{$^1$University of Rochester, Rochester, NY, USA \\
$^2$Carnegie Mellon University, Pittsburgh, PA, USA
$^3$Nagoya University, Nagoya, Japan
}

\begin{document}
\ninept
\maketitle
\begin{abstract}
With the advancements in singing voice generation and the growing presence of AI singers on media platforms, the inaugural Singing Voice Deepfake Detection (SVDD) Challenge aims to advance research in identifying AI-generated singing voices from authentic singers. This challenge features two tracks: a controlled setting track (CtrSVDD) and an in-the-wild scenario track (WildSVDD). The CtrSVDD track utilizes publicly available singing vocal data to generate deepfakes using state-of-the-art singing voice synthesis and conversion systems. Meanwhile, the WildSVDD track expands upon the existing SingFake dataset, which includes data sourced from popular user-generated content websites. For the CtrSVDD track, we received submissions from 47 teams, with 37 surpassing our baselines and the top team achieving a 1.65\% equal error rate. For the WildSVDD track, 
we benchmarked the baselines. This paper reviews these results, discusses key findings, and outlines future directions for SVDD research.



\end{abstract}
\section{Introduction}
\label{sec:intro}


The use of AI tools in music production has sparked considerable debate, introducing new issues and challenges~\cite{deruty2022development}. Notably, the development of advanced singing voice synthesis (SVS) and conversion (SVC) techniques has reached a significant milestone in AI-generated songs, where synthesized voices sound remarkably natural and align seamlessly with music scores. These synthesized voices can now emulate the vocal characteristics of any singer with minimal training data. While this technological advancement is impressive, it has raised widespread concerns among artists, record labels, and publishing houses~\cite{collins2024avoiding}. The potential for unauthorized synthetic reproductions that mimic well-known singers poses a real threat to the commercial value and intellectual property rights of original artists, prompting urgent calls for efficient and accurate methods to detect these deepfake singing voices.


In response to these concerns, research has emerged towards detecting AI-generated songs. Our prior research SingFake~\cite{zang2024singfake} has laid the groundwork for the emerging field of singing voice deepfake detection (SVDD). We introduced the SingFake dataset, a comprehensive collection of authentic and deepfake song clips featuring a variety of languages and singers. Concurrently, Xie et al.~\cite{xie2023fsd} curated FSD, a Chinese fake song detection dataset in a controlled setting. Both works found that speech-trained deepfake detection models experience significant performance degradation when tested on the SVDD task, highlighting the unique challenges brought by synthesized singing voices.


Our SingFake evaluations~\cite{zang2024singfake} highlighted several challenges, including dealing with unseen singers, various communication codecs, diverse languages and musical contexts, and interference from accompaniment tracks. This underscores the distinct nature of SVDD and the necessity for specialized SVDD systems. SingGraph~\cite{chen2024singing} has been a novel model that achieves state-of-the-art performance within the SingFake dataset. They utilized an acoustic understanding model and speech self-supervised learning (SSL) models, underscoring the synergy between music understanding and voice analysis for SVDD in the wild. A recent work~\cite{afchar2024detecting} proposed a general-purpose music deepfake detector that also includes generated instrumental parts, highlighting robustness and generalization issues.


To advance the field of SVDD, we introduce the SVDD challenge~\cite{zhang2024svdd}, the inaugural research initiative specifically dedicated to exploring SVDD. This challenge targets both controlled and in-the-wild settings, aiming to distinguish bonafide and AI-generated singing voice recordings within CtrSVDD and WildSVDD tracks, respectively. For the CtrSVDD track, we generated a dataset CtrSVDD~\cite{zang2024ctrsvdd} that exclusively uses clean, unaccompanied vocals provided by our data contributors, thereby mitigating the interference of voice separation algorithms. This two-track approach allows participants to tackle the challenges of identifying deepfake singing voices under different and realistic conditions. The WildSVDD track follows the same approach as our SingFake project~\cite{zang2024singfake}, dealing with deepfakes as they typically appear in online media, complete with background music. 


The SVDD challenge has received much attention, and the results have been promising. For the CtrSVDD track, we received submissions from 47 teams, with 37 surpassing our baselines and the top team achieving a 1.65\% equal error rate. Common strategies included self-supervised feature and ensemble models. For the WildSVDD track, we have not received any submissions yet, but we have benchmarked the baselines. Future research endeavors can focus on improving generalization ability with single systems and further exploring in-the-wild scenarios. We look forward to continued progress in this critical area of research.

\vspace{-3pt}

\section{Overview of challenge setups}

\vspace{-2pt}

\subsection{Two tracks: CtrSVDD and WildSVDD}

For singing voice deepfakes, these artificial creations are typically presented with background music, mirroring the authentic song experience. Our previous work on SingFake~\cite{zang2024singfake} followed this approach by collecting deepfakes from web videos. However, this method introduces a significant challenge: the process of separating vocals from music can create artifacts that obscure the differences between bonafide and deepfake vocals. 

To address the above issues comprehensively, the SVDD challenge is structured into two distinct tracks: \textit{Controlled} (CtrSVDD) and \textit{In-the-Wild} (WildSVDD) settings. The WildSVDD track mirrors the approach of the SingFake project, dealing with deepfakes as they typically appear in online media, including background instrumental interference and sometimes background noises. Conversely, the CtrSVDD track uses clean, unaccompanied vocals provided by our data contributors, thereby minimizing the interference caused by voice separation algorithms. The data construction process was described in~\cite{zang2024ctrsvdd}. We also train deepfake algorithms ourselves and thus can have all metadata of training configurations as well as source and target utterances. This double-track approach allows participants to tackle the challenges of identifying deepfake singing voices under both lab-controlled and realistic conditions.


For CtrSVDD, we generate singing vocals with existing SVS and SVC systems. The setting mitigates the artifacts by the singing voice separation algorithms, and we expect it to be easier than the in-the-wild settings. For WildSVDD, we continued the data collection process of SingFake to collect more data from the video platforms. This collection from the web inevitably introduces additional difficulties in the systematic analysis of SVDD systems due to unknown backend synthesis/conversion systems and post-processing strategies. The compression codecs used by the web platforms are also unknown and may change over time, causing inconsistencies across different downloads.


\subsection{Evaluation metric: EER}


We expect each SVDD system to generate a score file for every segmented clip, with higher scores indicating greater confidence that the clip is bonafide. The Equal Error Rate (EER) is achieved when the false acceptance and rejection rates are equal by varying the decision threshold, hence EER is independent of the threshold. A lower EER indicates a better system for distinguishing between bonafide and deepfake singing voices.

\section{CtrSVDD: Controlled singing voice deepfake detection}

\subsection{CtrSVDD database}
To construct the CtrSVDD database, we first collected bonafide samples from open-source singing recordings. These include\footnote{For comprehensive details of CtrSVDD, including links to source datasets and generation methods, please refer to~\cite{zang2024ctrsvdd}.}:
\vspace{-4pt}
\begin{itemize}
\item Mandarin datasets: Opencpop~\cite{wang2022opencpop}, M4Singer~\cite{zhang2022m4singer}, KiSing~\cite{shi2024singing}, and the official ACE-Studio release~\cite{timedomain2023acesinger}.
\vspace{-2pt}
\item Japanese datasets: Ofuton-P, Oniku Kurumi, Kiritan~\cite{ogawa2021tohoku}, and JVS-MuSiC~\cite{tamaru2020jvs}.
\end{itemize}
\vspace{-3pt}
We segmented these public dataset recordings into vocal clips. Next, we generated deepfake singing vocals using 14 existing SVS and SVC systems, as described in Table~\ref{tab:svs_svc_systems}. The SVS systems employed were ESPnet-Muskits~\cite{shi2022muskits}, NNSVS~\cite{yamamoto2023nnsvs}, DiffSinger~\cite{liu2022diffsinger}, and ACESinger~\cite{timedomain2023acesinger}. For SVC, we utilized NU-SVC~\cite{yamamoto2023comparative} and variants of So-VITS-SVC.
Specially for ACESinger, we asked the ACE Studio company to provide the bonafide utterances for a list of singers; we then generated deepfake data using the ACE Studio tool.

The CtrSVDD database is partitioned into training, development, and evaluation splits, following the methodology of the ASVspoof2019 speech anti-spoofing benchmark. The training and development sets use the same set of deepfake generation algorithms (A01-A08), while the evaluation set employs a different set (A09-A14). The database is released under a CC-BY-NC-ND 4.0 license, aligned with the sourcing corpora.
We released the training and development set\footnote{\url{https://zenodo.org/records/10467648}} and the evaluation set\footnote{\url{https://zenodo.org/records/12703261}}, while the keys and labels for the evaluation set were withheld until the submission deadline.

\begin{table}[t!]
    \centering
    \caption{Overview of the deepfake methods included in the CtrSVDD database. Please refer to~\cite{zang2024ctrsvdd} for detailed descriptions.}
\resizebox{\linewidth}{!}{
    \begin{tabular}{c|lcl}
        \toprule
        \multicolumn{1}{c|}{\textbf{System}} & \multicolumn{1}{c}{\textbf{Model}} & \multicolumn{1}{c}{\textbf{Type}} & \multicolumn{1}{c}{\textbf{Description}} \\
        \midrule
        A01  & XiaoiceSing        & SVS  & Cascaded Transformer model with a HiFi-GAN vocoder \\
        A02  & VISinger           & SVS  & End-to-end VAE with a HiFi-GAN vocoder \\
        A03  & VISinger2          & SVS  & End-to-end VAE with a DDSP vocoder \\
        A04  & NNSVS              & SVS  & Cascaded diffusion model with a source-filter HiFi-GAN \\
        A05  & Naive RNN          & SVS  & Cascaded LSTM model with a HiFi-GAN vocoder \\
        A06  & NU-SVC             & SVC  & NNSVS model with ContentVec linguistic features \\
        A07  & Soft-VITS-SVC      & SVC  & Soft-VITS model with WavLM linguistic features \\
        A08  & Soft-VITS-SVC      & SVC  & Soft-VITS model with ContentVec linguistic features \\
        \midrule
        A09  & Soft-VITS-SVC      & SVC  & Soft-VITS model with additional source-filter HiFi-GAN \\
        A10  & Soft-VITS-SVC      & SVC  & Soft-VITS model with MR-HuBERT linguistic features \\
        A11  & Soft-VITS-SVC      & SVC  & Soft-VITS model with WavLabLM linguistic features \\
        A12  & DiffSinger         & SVS  & Cascaded Transformer model with a post diffusion module  \\
        A13  & Soft-VITS-SVC      & SVC  & Soft-VITS model with Chinese HuBERT linguistic features \\
        A14  & ACESinger          & SVS  & Blackbox commercial system with manual tuning \\
        \bottomrule
    \end{tabular}
    }
    \label{tab:svs_svc_systems}
    \vspace{-10pt}
\end{table}

\begin{table}[t]
\centering
\renewcommand{\arraystretch}{0.9}
\caption{Summary of the CtrSVDD database~\cite{zang2024ctrsvdd}.
}
\label{table: CtrSVDD}
\resizebox{0.8\columnwidth}{!}{
\begin{tabular}{@{}lcccc@{}}
\toprule
\multirow{2}{*}{Partition} & \multirow{2}{*}{\begin{tabular}[c]{@{}c@{}}\# Singers\end{tabular}}  
& Bonafide     
& \multicolumn{2}{c}{Deepfake} \\ 
\cmidrule(l){3-5}
& &\# Clips & \# Clips   
& Methods     \\ 
\midrule
Training   & 59      & 12,169  & 72,235 & A01$\sim$A08     \\
Development & 55     & ~~6,547  & 37,078 & A01$\sim$A08     \\
Evaluation  & 48     & 13,596  & 79,173 & A09$\sim$A14     \\ 
\bottomrule
\end{tabular}}
\vspace{-15pt}
\end{table}

\subsection{Protocols}
 
Note that the training and development sets available on Zenodo are incomplete because of licensing issues with some bonafide datasets. To fully retrieve the dataset, we asked the participants to first download all the remaining bonafide datasets on their own by agreeing to the terms and conditions and then follow the instructions we provided\footnote{\url{https://github.com/SVDDChallenge/CtrSVDD_Utils}}. Participants can refer to the statistics in Table~\ref{table: CtrSVDD} as a guide to verify the completion of their downloads and generation.



Participants are asked to ask teams to score each vocal clip. Using the submitted scores, we calculate and rank participant systems using EER. 
During our baseline analysis~\cite{zang2024ctrsvdd}, we found the A14 attack (originated from ACESinger) shows strong out-of-domain attributes~\cite{zang2024ctrsvdd}. As such, we exclude the A14 attack and the bonafide utterances of the ACESinger during the official ranking of the CtrSVDD track.
Firstly, unlike other deepfake attacks (A09-A13) that were trained using open-source datasets, the A14 attack was trained on unknown data sources. Additionally, the bonafide portion of ACESinger was not used to train the other attack systems (A09-A13). The bonafide portion of ACESinger was provided by Timedomain, a company that may apply additional processing steps to the audio output, while the deepfake portion was generated using their ACE Studio desktop application.


\subsection{Submission rules}
We used CodaBench~\cite{codalab_competitions_JMLR} for CtrSVDD results submission. Each team is allowed a maximum of three submissions for the entire duration of the CtrSVDD challenge for official ranking purposes. This limit is in place to ensure fairness and to encourage strategic submissions. It is important to note that CodaBench's daily submission limit is separate; our three-submission limit refers to the total allowable submissions for the challenge. After the three initial submissions are used, participants may utilize an additional CodaBench dedicated to research. This allows them to submit scores and obtain per-attack, per-dataset, and overall EERs for research purposes.


Participants are welcome to use any publicly available datasets for training in addition to the CtrSVDD we provide, but of course, exclude any datasets used in our test set. Specifically, for the CtrSVDD track, participants must not use M4Singer, KiSing, any open-sourced deepfake models based on M4Singer and/or KiSing, or the commercial software ACE Studio~\cite{timedomain2023acesinger}.
We refer the participants to the list of available datasets in the evaluation plan~\cite{zhang2024svdd}.


\subsection{Baseline solutions}

We have developed two baseline systems for the challenge: one that uses raw waveforms and another that employs linear frequency cepstral coefficients (LFCCs) as front-end features:
The raw waveform system is an AASIST~\cite{Jung2021AASIST}-based system.
The LFCC system uses 60 coefficients, with a 512 sample window and 160 sample hop size. The LFCC features pass through several downsampling residual convolution blocks and a linear layer connecting it to the graph attention network backend of~\cite{Jung2021AASIST}. 

We refer to the LFCC system as \textbf{B01} and the raw waveform model as \textbf{B02}. For both systems, we conducted training over 100 epochs using a fixed random seed, exclusively on the CtrSVDD training partition. We then selected the checkpoint that achieved the lowest validation EER on the CtrSVDD development partition for evaluation.
During training and evaluation, the models processed 4-second random audio segments from each utterance. Details of the implementation are available on the challenge GitHub repository\footnote{\url{https://github.com/SVDDChallenge/CtrSVDD2024_Baseline}}.

\section{WildSVDD: In-the-wild singing voice deepfake detection}

\subsection{WildSVDD database}

We gathered data annotations from social media platforms following a method similar to the SingFake project~\cite{zang2024singfake}. The WildSVDD dataset has been expanded to approximately double the original size of SingFake, now featuring 97 singers with 2007 deepfake and 1216 bonafide songs. The annotators, who were familiar with the singers they covered, manually verified the user-specified labels during the annotation process to ensure accuracy, especially in cases where the singer(s) did not actually perform certain songs. 
We cross-checked the annotations against song titles and descriptions, and manually reviewed any discrepancies for further verification.
We verified the accessibility of all URLs in the dataset as of March 28th and removed any that were inaccessible. The WildSVDD dataset now includes Korean singers, making Korean the third most represented language in the dataset. To help track changes between the SingFake and WildSVDD datasets, we have added a "SingFake\_Set" column that indicates the original partition of an annotation in the SingFake dataset. Annotations that lack a value in this column are new additions to the WildSVDD dataset. We form two test sets: Test A denotes a newly formed testing dataset including new samples, while Test B denotes the hardest test set T04 detailed in~\cite{zang2024singfake}.





Due to potential copyright issues, we only released the annotations \footnote{\url{https://zenodo.org/records/10893604}} under a CC-BY 4.0 license. Consequently, participants might acquire slightly different media files that correspond to the same annotations, depending on the specifics of their download process. Due to this variability, self-reported metrics from participants can, at best, be used as a rough reference and cannot be directly used to compare systems.
As such, we encourage participants to report the success rate of URL downloads per partition and, if possible, the actual files used for training and testing. This transparency allows researchers to make fairer comparisons. Additionally, participants are encouraged to open source their model implementations to facilitate the reproduction of results with the WildSVDD dataset.

\subsection{Protocols}

We provide the training and test partitions, allowing participants the flexibility to carve out a validation set from the training data for model development. We provide labels of SingFake~\cite{zang2024singfake} partitions for annotations that appeared in the SingFake dataset for easy comparison with previous systems.
The test set is divided into parts A and B, with part B considered more challenging due to its inclusion of unseen musical contexts~\cite{zang2024singfake}.

We recommend that participants further segment the songs into clips using our tool available in the SingFake GitHub repository\footnote{\url{https://github.com/yongyizang/SingFake}}. Evaluations should be conducted at the segment level rather than at the song level.
We adopt the self-reported EER and do not accurately rank the results. We encourage the participants to submit the score files listing the URLs, segment start and end timestamps, and the corresponding scores output from their systems.


\subsection{Baseline solutions}
To establish baseline solutions, we implement the architecture described in~\cite{zang2024ctrsvdd}, applying both mixtures and vocals configurations as outlined in~\cite{zang2024singfake}. We also incorporate the self-supervised learning model XLS-R~\cite{babu2021xls} based system proposed in~\cite{tak2022automatic}, given its popularity among top performers in the CtrSVDD track. Existing research and CtrSVDD track outcomes indicate that data augmentation significantly enhances performance with self-supervised frontends. Consequently, we present results both with and without augmentation. 
We adhere to the training schedule, learning rate, and optimizer specified in~\cite{zang2024ctrsvdd}, with the exception of XLS-R based models, where we adopt training settings same as~\cite{tak2022automatic}. 

\begin{table}[]
\centering
\renewcommand{\arraystretch}{0.9}
\caption{Comparison of EERs of different baseline front-end processing methods and settings on WildSVDD track.}
\resizebox{0.9\linewidth}{!}{
\begin{tabular}{lcccc}
\toprule
\multirow{2}{*}{Front-end} & \multicolumn{2}{c}{WildSVDD Test A} & \multicolumn{2}{c}{WildSVDD Test B} \\
\cmidrule(lr){2-3} \cmidrule(lr){4-5}
 & Mixtures & Vocals & Mixtures & Vocals \\
\midrule
Raw Waveform & 10.50 & ~~8.48 & 16.85 & 14.91 \\
Spectrogram & 27.93 & 20.55 & 30.97 & 24.41 \\
Mel-Spectrogram & 29.27 & 27.35 & 32.18 & 30.78 \\
MFCC & 17.78 & 19.14 & 22.92 & 23.31 \\
LFCC & 22.60 & 23.25 & 26.82 & 26.94 \\
Wav2vec2 XLS-R & ~~9.57 & ~~6.09 & 21.45 & 24.09 \\
\bottomrule
\end{tabular}
}
\label{tab:front-end-comparison}
\vspace{-15pt}
\end{table}

Notably, we observe that the raw waveform system's performance on WildSVDD Test B, which is equivalent to SingFake T04, shows significant improvement compared to~\cite{zang2024singfake}, despite highly similar system architectures. This improvement suggests that data-centric approaches may prove most effective in the long run. The most substantial difference between these two system versions lies in the volume of training data available, highlighting the potential impact of increased data resources on model performance. 



\section{Challenge results}


We received 89 registration forms, with 84 teams expressing interest in participating in the CtrSVDD track and 65 teams showing interest in the WildSVDD track. Out of these, 47 teams submitted their results to the CtrSVDD track. Unfortunately, we have not received any submissions for the WildSVDD track, probably due to time constraints and copyright concerns.

\subsection{CtrSVDD results from the team submissions}
 
Table~\ref{tab:ctrsvdd_res} presents the results from the CtrSVDD track of the SVDD challenge, ranking the participating teams based on their EER without ACESinger and providing additional EER values for all attacks. Most teams used their three allowed entries for submission, while some did not but still achieved commendable ranks. We ranked the teams based on the best submission (lowest EER) among their entries. Among the 47 submissions (excluding two baselines), 37 surpassed both baselines, demonstrating substantial progress in developing SVDD systems capable of distinguishing between bonafide and deepfake singing voices.

\begin{table}[t!]
\renewcommand{\arraystretch}{0.9}
\centering
\caption{Summary of the CtrSVDD challenge results. The EER without ACESinger is used as the evaluation metric to rank the submissions, while the EER for all attacks is listed for analysis. The rows for both baseline systems are shaded. Teams with bolded ranks submitted the system description. }
\resizebox{\linewidth}{!}{
\begin{tabular}{c|lcc|r}
\toprule
\multicolumn{1}{c|}{\textbf{Rank \#}} & \multicolumn{1}{c}{\textbf{Team Name}} & \multicolumn{1}{c}{\textbf{\# Entries}} & \multicolumn{1}{c}{\textbf{\begin{tabular}[c]{@{}c@{}}EER \\ (w/o ACESinger)\end{tabular}}} & \multicolumn{1}{|c}{\textbf{\begin{tabular}[c]{@{}c@{}}EER \\ (overall)\end{tabular}}} \\
\midrule
~~\textbf{1}                                    & Fosafer Speech                         & 3                                       & ~~1.65                                                                                  & 4.32~~~                                                                              \\

~~\textbf{2}                                    & NBU\_MISL                              & 3                                       & ~~2.00                                                                                  & 8.41~~~                                                                              \\
~~\textbf{3}                                    & I2R-ASTAR                              & 3                                       & ~~2.22                                                                                  & 4.86~~~                                                                              \\
~~\textbf{4}                                    & Qishan                                 & 2                                       & ~~2.32                                                                                  & 4.45~~~                                                                              \\
~~5                                    & Breast waves                           & 3                                       & ~~2.73                                                                                  & 5.38~~~                                                                              \\
~~6                                    & MediaForensics                         & 3                                       & ~~2.75                                                                                  & 5.83~~~                                                                              \\
~~7                                    & beyond                                 & 3                                       & ~~2.99                                                                                  & 5.68~~~                                                                              \\
~~8                                    & Star                                   & 2                                       & ~~3.31                                                                                 & 5.21~~~                                                                              \\
~~9                                    & shrinep                                & 3                                       & ~~3.53                                                                                  & 5.91~~~                                                                              \\
10                                   & HUBENMINZU                             & 3                                       & ~~3.61                                                                                  & 5.63~~~                                                                              \\
\textbf{11}                                   & Pindrop                                & 3                                       & ~~3.85                                                                                  & 6.27~~~                                                                              \\
12                                   & SVDD-Xin                               & 2                                       & ~~4.05                                                                                  & 8.17~~~                                                                              \\

13                                   & DDD                                    & 3                                       & ~~5.33                                                                                  & 7.57~~~                                                                              \\
14                                   & rudraprasad                            & 2                                       & ~~5.45                                                                                  & 9.87~~~                                                                             \\
15                                   & beautifulboy                           & 3                                       & ~~5.79                                                                                  & 7.50~~~                                                                              \\
16                                   & WinterIsComing                         & 2                                       & ~~6.71                                                                                  & 7.37~~~                                                                              \\
17                                   & bokingchen                             & 3                                       & ~~6.76                                                                                  & 17.98~~~                                                                             \\
18                                   & UCAS\_2024                             & 2                                       & ~~6.81                                                                                  & 7.48~~~                                                                              \\
18                                   & ColdLightXXX                           & 3                                       & ~~6.81                                                                                  & 9.97~~~                                                                              \\
\textbf{20}                                   & xieyuankun                             & 1                                       & ~~6.90                                                                                & 7.95~~~                                                                              \\
21                                   & xyyuan                                 & 3                                       & ~~7.36                                                                               & 8.35~~~                                                                              \\
22                                   & tarrifin                               & 3                                       & ~~7.56                                                                                  & 17.62~~~                                                                             \\
23                                   & LLLSLin                                & 3                                       & ~~7.90                                                                                  & 8.27~~~                                                                              \\
24                                   & EEGBrain                               & 3                                       & ~~7.92                                                                                  & 10.46~~~                                                                             \\
25                                   & MelodyAI                               & 1                                       & ~~8.66                                                                                  & 9.66~~~                                                                              \\
26                                   & Outlaw Monkeys                         & 1                                       & ~~8.73                                                                                  & 13.24~~~                                                                             \\

27                                  & starbucks                              & 3                                       & ~~8.82                                                                                  & 12.69~~~                                                                             \\
28                                   & ForgeryMark                            & 3                                       & ~~8.85                                                                                  & 8.84~~~                                                                              \\
29                                   & test\_acc                              & 3                                       & ~~8.87                                                                                 & 14.53~~~                                                                             \\
\textbf{30}                                   & notepad                                & 3                                       & ~~8.99                                                                                  & 11.88~~~                                                                             \\
31                                   & Feathers                               & 2                                       & ~~9.22                                                                                  & 14.32~~~                                                                             \\

32                                   & asada                                  & 1                                       & ~~9.64                                                                                  & 10.66~~~                                                                             \\
33                                   & fengchuippshuang                       & 1                                       & 10.14                                                                                  & 9.91~~~                                                                              \\
34                                   & USC                                    & 2                                       & 10.31                                                                                  & 13.05~~~                                                                             \\
35                                   & JAIST                                  & 2                                       & 10.39                                                                                  & 12.68~~~                                                                             \\
36                                   & Polimi-Unibs                           & 3                                       & 10.50                                                                                 & 22.58~~~                                                                             \\
37                                   & liuziyi                                & 3                                       & 10.80                                                                                 & 19.76~~~                                                                             \\
\rowcolor{gray!50}38                                   & B02                                    &                                        & 11.16                                                                                 & 13.58~~~                                                                             \\
38                                   & jiaruiliu                              & 1                                       & 11.16                                                                                 & 14.70~~~                                                                             \\
40                                   & SynthSound                             & 3                                       & 11.21                                                                                 & 14.54~~~                                                                             \\
41                                   & zzgww                                  & 3                                       & 11.27                                                                                 & 12.60~~~                                                                             \\
42                                   & AIS                                    & 3                                       & 11.67                                                                                 & 15.52~~~                                                                             \\
\rowcolor{gray!50}43                                   & B01                                    &                                        & 12.03                                                                                 & 16.10~~~                                                                             \\
44                                   & B401                                   & 3                                       & 12.39                                                                                 & 15.92~~~                                                                             \\
45                                   & datajedi23                             & 2                                       & 13.50                                                                               & 18.32~~~                                                                             \\
46                                   & Zhejiang University                    & 3                                       & 15.91                                                                                 & 18.48~~~                                                                             \\
\textbf{47}                                   & vitas                                  & 3                                       & 16.05                                                                                 & 28.10~~~                                                                             \\
48                                   & Dashlab                                & 3                                       & 17.04                                                                                 & 27.13~~~         \\
49                                   & jiachengdeng                           & 3                                       & 17.44                                                                                 & 20.45~~~                                                                             \\

\bottomrule
\end{tabular}
}
\label{tab:ctrsvdd_res}
\vspace{-15pt}
\end{table}

\begin{table*}[t!]
\centering
\renewcommand{\arraystretch}{0.9}
\caption{Overview of the top-8 ranked submission results. The best performance in each column is \textbf{bolded} and the second is \underline{underlined}.}
\label{tab:subtract-results}
\resizebox{0.92\textwidth}{!}{%
\begin{tabular}{l|cccc|ccccccc|c}
\toprule
\multicolumn{1}{c|}{\multirow{2}{*}{\textbf{Team Name}}} & \multicolumn{2}{c}{Results (w/o ACESinger)} & \multicolumn{2}{c|}{Results (overall)} & \multicolumn{5}{c}{Per-Attack EER}                                            & \multicolumn{2}{c|}{Per-Dataset EER} & \multirow{2}{*}{ACESinger (A14)} \\ 
\cmidrule(lr){2-3} \cmidrule(lr){4-5} \cmidrule(lr){6-10} \cmidrule(lr){11-12}
\multicolumn{1}{c|}{}                           & EER (\%)               & Rank               & EER (\%)             & Rank            & A09           & A10           & A11           & A12           & A13           & KiSing        & M4Singer             &                                  \\
\midrule
Fosafer Speech                                  & \textbf{1.65}                   & 1                  & \textbf{4.32}                 & ~~1               & \underline{0.23}          & \textbf{0.06} & \textbf{0.37} & \textbf{4.19} & \textbf{0.07} & \underline{2.66}          & \textbf{1.69}        & \underline{49.67}                            \\
NBU\_MISL                                       & \underline{2.00}                   & 2                  & 8.41                 & 19              & \textbf{0.13} & \underline{0.11}          & \underline{0.94}          & 5.17          & \underline{0.10}          & 8.98          & \underline{2.07}                 & 50.02                            \\
I2R-ASTAR                                       & 2.22                   & 3                  & 4.86                 & ~~3               & 0.65          & 0.51          & 2.49          & 4.57          & 0.64          & 6.01          & 2.16                 & 50.02                            \\
Qishan                                          & 2.32                   & 4                  & \underline{4.45}                 & ~~2               & 1.02          & 0.69          & 2.54          & 4.42          & 0.76          & 2.82          & 2.32                 & 50.05                            \\
Breast waves                                    & 2.73                   & 5                  & 5.38                 & ~~5               & 1.50          & 0.76          & 2.03          & 6.14          & 0.88          & 3.56          & 2.84                 & 50.44                            \\
MediaForensics                                  & 2.75                   & 6                  & 5.83                 & ~~8               & 0.56          & 0.38          & 3.90          & 4.45          & 1.02          & 10.56         & 2.56                 & 49.91                            \\
beyond                                          & 2.99                   & 7                  & 5.68                 & ~~7               & 0.45          & 0.26          & 4.56          & \underline{4.37}          & 0.85          & 9.12          & 2.85                 & \textbf{49.53}                   \\
Star                                            & 3.31                   & 8                  & 5.21                 & ~~4               & 1.64          & 0.19          & 1.11          & 7.30          & 0.23          & \textbf{1.79}          & 3.51                 & 49.70                           \\
\bottomrule
\end{tabular}%
}
\vspace{-15pt}
\end{table*}


The baseline systems B01 and B02 achieved EERs of 12.03\% and 11.16\%, respectively. The raw-waveform-based system outperformed the LFCC-based system, which was consistent with observations in the speech anti-spoofing task. The top-ranked team achieved an impressive EER of 1.65\%, significantly lower than the baselines.
The top teams' results were very close, highlighting the competitive nature of the challenge. However, several teams' performances fell between or even below the baseline results, underscoring the challenge's difficulty. These results emphasize the complexity of SVDD, especially when considering diverse unseen attack methods. 


The overall EERs that include the ACESinger bonafide portion and A14 attack are consistently higher than those that exclude them, indicating the difficulty of detecting deepfakes generated from out-of-domain commercial SVS systems. While the trend of increasing EERs is generally similar, some systems achieved much lower or higher overall EERs than expected. This leads to our detailed analysis of per-attack EER and per-dataset EER in the next subsection.



\subsection{Per-attack and per-dataset analysis}

For the per-dataset EER, we calculate the EER based on the metadata of the source bonafide datasets, specifically identifying which dataset the target speaker belongs to. For the per-attack EER, A09-A13 is calculated using all bonafide data except ACESinger, along with the deepfake data generated by the corresponding methods. The A14 EER is calculated using the ACESinger bonafide portion and the deepfakes generated by the ACE Studio tool, making it identical to the per-dataset EER for ACESinger.
Note that we used a slightly different metric during the challenge but changed post-challenge. Previously, per-attack EERs on A09-A14 were calculated using all bonafide song clips against deepfake clips for each category~\cite{zang2024ctrsvdd}, which could be what was reported by participants.

Table~\ref{tab:subtract-results} provides a detailed analysis of the top eight ranked submissions. The per-attack results do not show a consistent decline across all attacks as rankings improve. Notably, the performance on A11 and A12 suggests that the top-performing systems may not necessarily be more robust to different types of attacks. This observation is consistent with findings in speech anti-spoofing research~\cite{nautsch2021asvspoof}.
Besides, The systems' performance on A12 is generally worse compared to other attacks, which is consistent with our baseline analysis~\cite{zang2024ctrsvdd}.
Furthermore, although the top four systems achieved similar EERs, their robustness differs significantly. The ``NBU\_MISL'' system is less robust to KiSing and ACESinger compared to other top systems, dropping to 19th place when ACESinger is factored in.

A similar phenomenon is evident in the per-dataset results. The best performance on KiSing was achieved by the team ``Star'', which was ranked 8th. While KiSing and M4Singer datasets yield overall low EERs, ACESinger consistently produced EERs around 50\%, approximating random guessing. 
The performance gap observed with ACESinger is likely due to inconsistencies between ACESinger and the rest of the CtrSVDD dataset. 
Notably, two teams achieved 93\% and 84\% on ACESinger, indicating some ability to establish a decision boundary on this dataset, albeit with incorrect decisions. 

Both the per-attack and per-dataset EERs highlight the ongoing challenge of generalization to unseen generation methods and acoustic channels. This underscores the need for increasing research efforts focused on developing systems that are more robust to these out-of-distribution conditions.


\subsection{Solution strategies}

\vspace{-1pt}

\begin{figure}[t!]
    \centering
    \includegraphics[width=0.99\linewidth]{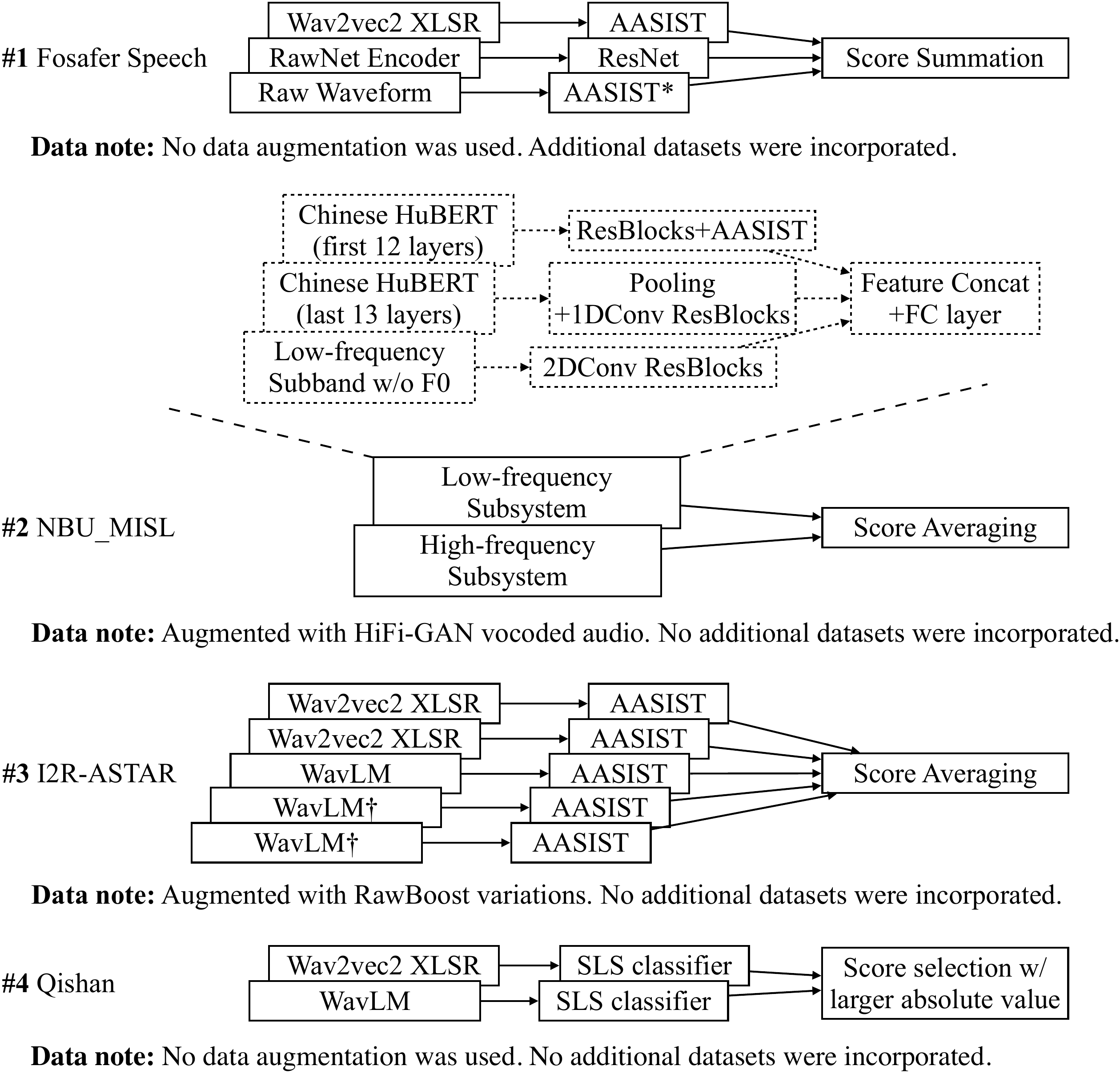}
    \caption{Illustration of the strategies employed by the top-4 ranked system submissions for the CtrSVDD track. An asterisk (*) indicates the additional use of adversarial training strategies for AASIST. A dagger (\dag) denotes different layer aggregation strategies proposed for WavLM, as opposed to the weighted sum method.}
    \label{fig:sys_desc}
    \vspace{-12pt}
\end{figure}


Among all submissions, 8 teams submitted system descriptions, with their ranks bolded in Table~\ref{tab:ctrsvdd_res}. Based on the submitted strategies, most teams utilized self-supervised learning (SSL) frontends and ensemble learning. For features, both raw waveform and SSL features were extensively explored. The most popular SSL feature used is wav2vec2 XLSR~\cite{babu2021xls}, a cross-lingual representation. Popular backend choices included ResNet and AASIST~\cite{Jung2021AASIST}, while score averaging was the favored ensemble method.


All the top four teams share similar strategies. Their general schemes are illustrated in Figure~\ref{fig:sys_desc}. The top-ranked team, ``Fosafer Speech,'' used additional datasets besides CtrSVDD: HiFi-TTS~\cite{bakhturina2021hi}, OpenSinger~\cite{huang2021multi}, CSD~\cite{choi2020children},  itako-Singing,
JSUT-Song,
Namine\_ritsu\_utagoe\_db,
no7-singing,
PJS~\cite{koguchi2020pjs}, PoPCS~\cite{liu2022diffsinger}, URSing~\cite{Li-2021}, which are all among the list of allowable external training datasets in our evaluation plan~\cite{zhang2024svdd}. They also generated additional training data, nhv\_share, and made it public\footnote{\url{https://drive.google.com/file/d/1h36C6mWvywIYXSPErDf2tcYfPuRkGKoQ/}} \footnote{\url{https://drive.google.com/file/d/1LhbH2-yNe_ZHcmTxqS5q7XQYlL1_JM_-/}}, one week before the challenge submission deadline as requested. They fused three subsystems and incorporated adversarial training~\cite{ganin2015unsupervised} to improve domain adaptation in the third system. 
The team ``NBU\_MISL'' developed two subsystems focused on low and high frequencies, respectively. Both used feature aggregation from Chinese HuBERT and the corresponding frequency band. They augmented the dataset with HiFi-GAN~\cite{kong2020hifi} vocoded audio.
The team ``I2R-ASTAR'' developed five subsystems with various layer aggregation strategies~\cite{pan2024attentive} and variants of the RawBoost data augmentation method for each subsystem. Please refer to~\cite{guragain2024speech} for more details of their system.
The team ``Qishan'' developed two subsystems with different SSL features. Each subsystem follows a Sensitive Layer Select (SLS) classifier that uses an adaptive weight allocation method~\cite{zhang2024audio} to aggregate SSL features and pool the feature map to a score. The score with a larger absolute value is selected for submission.


The team ``Pindrop'' developed three subsystems based on x-vector, ResNetSE34, and wav2vec2 XLSR. Additional datasets were incorporated. The team ``xieyuankun'' submitted a single system using wav2vec2 XLSR combined with AASIST, without any data augmentation. This approach could serve as an addition to our baseline systems. 
More systems have been submitted to the research track of CodaBench, some of which demonstrate better performance than their challenge submissions. We look forward to paper submissions to learn more about their strategies.



\section{Discussions and conclusions}


The CtrSVDD track of the SVDD challenge was a notable success, attracting 47 submissions, with 37 surpassing the baseline performance. The top teams employed diverse and advanced techniques, such as self-supervised learning, ensemble learning, and adversarial training, demonstrating significant innovation in the field. Detailed system descriptions from eight teams provided valuable insights for future research. This success highlights the progress in deepfake detection for singing voices and sets the stage for further advancements and improvements.


The lack of submissions for the WildSVDD track can be attributed to concerns over copyright issues, the complexities of the download process, and the time-consuming nature of data preparation. Additionally, challenges with the reproducibility of baseline methods and the limited time frame may have discouraged participation. Teams might have prioritized the CtrSVDD track, where their efforts would be more recognized and impactful.




For future work, the CtrSVDD track can be enhanced by including the latest advancements in SVS and SVC techniques. 
Additionally, cross-database evaluations between the CtrSVDD and WildSVDD datasets could be conducted to assess the generalizability and robustness of SVDD systems, offering an intriguing research avenue. Exploring a broader variety of singing deepfake types is also recommended. Given recent advancements in singing generative models that can produce not only vocals but also accompanying music, as noted in~\cite{donahue2023singsong}, it becomes crucial to develop detection systems capable of identifying AI-generated songs in their entirety.

\vfill\pagebreak
\footnotesize
\bibliographystyle{IEEEbib}
\bibliography{refs}
\end{document}